\documentclass{basi}

\usepackage{graphics}
\usepackage{graphicx}
\usepackage{epsfig}
\include{bk_macros}
\def\bfig{\begin{figure}}
\def\efig{\end{figure}}
\def\bcenter{\begin{center}}
\def\ecenter{\end{center}}

\begin{document}
\title{A Critique of the relativistic cosmology}

\author[R. K. Thakur]
       {R. K. Thakur,\thanks{e-mail:rkthakur0516@yahoo.com}\thanks{21 College Road, Choubey Colony, Raipur - 492001, India.}\\
        Rtd. Professor of Physics, School of Studies in Physics,\\Pt. Ravishankar Shukla University, Raipur-India.} 
\maketitle
\label{firstpage}
\begin{abstract}
All the relativistic cosmological models of the universe, except Einstein's static model, imply that the 3-space of the spacetime of the universe is also expanding apart from the matter and the radiation in it. However, there is no observational evidence of the expansion of the 3-space of the spacetime of the universe. Actually, the 3-space of the spacetime of the universe might not be expanding at all. Consequently, the conceptual foundation of the relativistic cosmological models of the universe based on the general theory of relativity, which in turn is based on Riemannian geometry, might be faulty and misleading.
\end{abstract}

\section{Introduction}
Relativistic cosmological models of the universe are based on either Einstein's field equations
\begin{eqnarray}
R_{\mu\nu} - \frac{1}{2}Rg_{\mu\nu} = -\kappa T_{\mu\nu}
\end{eqnarray}
of the general theory of relativity (GTR) published in 1915 (Einstein 1915), or his modified field equations
\begin{eqnarray}
R_{\mu\nu} - \frac{1}{2}Rg_{\mu\nu} + \Lambda g_{\mu\nu} = -\kappa T_{\mu\nu}
\end{eqnarray}
published in 1917 (Einstein 1917), where $\mu,\nu$ run over the four values 0, 1, 2, 3.
In eqs.(1) and (2) $g_{\mu\nu}$, the ten functions of the coordinates $x^{\circ} = ct,~x^{1},~x^{2},~x^{3}$ are the components of the fundamental symmetric metric tensor, $R_{\mu\nu}$ that of the Ricci tensor, $ R= g^{\mu\nu}R_{\mu\nu}$ (the summation over the repeated indices is implied here and in the sequel) the scaler curvature, $ \kappa = 8\pi G/c^4$ (where G is Newton's constant of gravitation, and c the speed of light in vacuum), $T_{\mu\nu}$ the components of the energy-momentum tensor, and $\Lambda$ the cosmological constant. The term $\Lambda g_{\mu\nu}$ in eqs.(2) is called the `cosmological term'.

In order to solve the `cosmological problem' Einstein assumed that on a large scale, the universe is homogeneous and isotropic, an assumption later enshrined as the 'cosmological principle' by cosmologists. Using this assumption and the `dust' approximation (wherein it is assumed that the random velocities of the `particles' of matter in the universe are zero), when Einstein tried to solve the field eqs.(1), to his dismay he could not obtain a static solution. At that time Einstein was of the view that the universe was static and had so much matter as to `close' it. Therefore, in order to obtain a static solution, in 1917 he modified his field eqs.(1) to (2). Consequently, he did obtain a static solution of his eqs.(2) for $\Lambda > 0 $. The corresponding model of the universe is known as the `Einstein universe'. The line element $ds$ in the Einstein universe is given by
\begin{eqnarray}
ds^2 = c^{2}dt^{2} - S^{2}\left[\frac{dr^2}{1-r^2} + r^2(d\theta^2 + sin^2\theta d\phi^2) \right]
\end{eqnarray}
in $r,~\theta,~\phi$ coordinates. In eq.(3) $S$ is a constant, and is called the `radius of the universe.'

Einstein could show that for positive value of $\Lambda$ eqs.(2) admitted of a solution in which the density of matter in the universe was uniform, its random velocities zero, and in which the space was so curved that although unbounded, it was finite. Furthermore, he thought that for positive value of $\Lambda$ eqs.(2) had no solution for $T_{\mu\nu} = 0$, i.e., for empty universe. Einstein was of the opinion that these points implied that his much cherished Mach's principle  had been incorporated into his GTR.

\section{Expanding Universe}
 However, shortly after the publication of Einstein's paper, in 1917 itself, W. de Sitter published a solution of the field eqs.(2) for empty space for positive value of $\Lambda$ (de Sitter1917a,b,c,d). This solution, known as the `de Sitter universe' is not static, it implies that the universe is `expanding'. Consequently, the pertinent question is : What is expanding in the de Sitter universe ? The obvious answer is : The 3-space is expanding in the de Sitter universe. If so, then the next question is : Why is the de Sitter universe expanding and what triggered off the expansion ?

Another fall out of the `discovery' of the de Sitter universe is that it proved that Einstein was wrong in asserting that eqs.(2) had no solution for positive value of $\Lambda$ for $T_{\mu\nu} = 0$, and as such in claiming that Mach's principle had been incorporated into his GTR. As a result of this Einstein abandoned the cosmological term and with it the attempt to incorporate Mach's principle into his GTR(Bondi 1948). After Hubble's announcement in 1929 of an apparent proportionality between the redshift and the distance of galaxies (Hubble 1929), which was the culmination of the earlier work of Slipher (1914, 1924), Wirtz (1921, 1922, 1924, 1925), Lundmark (1920, 1924, 1925), and Hubble himself, that led to the concept of an `expanding' universe, in a private conversation, Einstein remarked to George Gamow that the introduction of the cosmological term was the biggest blunder he ever made in his life (Gamow 1970). Nevertheless, Lema{\^{\i}}tre has forcefully argued in favour of the introduction of the cosmological term (Lemaitre 1970) to which Einstein's rebuttal is not unreasonable (Einstein 1970). However, in recent years the cosmological constant is being resurrected after the discovery of the `dark energy' which is thought to be the manifestation of the cosmological constant (Peebles \& Ratra 2003).

\section{Implied expansion of the 3-space}
Except Einstein's static model of the universe (i.e., the Einstein universe) all the relativistic cosmological models of the universe imply that the universe is `expanding'. They imply that not only the matter and the radiation in the universe are expanding, but also the 3-space of the spacetime (i.e., what is known as the `space' in common parlance) of the universe is also expanding. This is obvious from the perusal of the following expressions for the line element in some of the well-known relativistic cosmological models of the universe :
\begin{enumerate}
\item
In the de Sitter model of the universe based on eqs.(2) the line element is given by
\begin{eqnarray}
ds^2 = c^2 dt^2 - e^{2Ht} \left[dr^2 + r^2(d\theta^2 + sin^2\theta d\phi^2) \right]
\end{eqnarray}
where H is a constant related to the cosmological constant $\Lambda$ through the equation
\begin{eqnarray}
H^2 = \frac{\Lambda c^2}{3}
\end{eqnarray}
\item
In the Friedmann models of the universe (Friedmann 1922, 1924) based on eqs.(1) the line elements can be written in the Robertson-Walker form (Robertson 1935; Walker 1936)
\begin{eqnarray}
ds^2 = c^2 dt^2 - S^2(t)\left[\frac{dr^2}{1-kr^2} + r^2(d\theta^2 + sin^2\theta d\phi^2) \right]
\end{eqnarray}
where the scale-factor (also called the expansion-factor) S(t) is a function of the cosmic time t and k=1, 0 or -1 according as the 3-space of the spacetime of the universe is positively curved, flat or negatively curved.
\item
In the Lema{\^{\i}}tre models of the universe (Lemaitre 1927, 1931) based on eqs.(2) the line element is taken to be 
\begin{eqnarray}
ds^2 = c^2 dt^2 - S^2(t)\left[\frac{dr^2}{1-r^2} + r^2(d\theta^2 + sin^2\theta d\phi^2) \right]
\end{eqnarray}
\item
In the steady-state model of the universe (Bondi \& Gold 1948; Hoyle 1948) based on the perfect cosmological principle (which states that in addition to the homogeneity and isotropy of the universe implied in the cosmological principle, the universe is homogeneous in time also, i.e., on a large scale, it is unchanging with time) the line element is taken to be
\begin{eqnarray}
ds^2 = c^2 dt^2 - e^{2H_\circ t}\left[dr^2 + r^2(d\theta^2 + sin^2\theta d\phi^2) \right]
\end{eqnarray}
In this model the expansion factor $a(t) = e^{H_\circ t}$, and the Hubble constant
\begin{eqnarray}
H = \frac{\dot{a}}{a} = constant = H_\circ
\end{eqnarray}
i.e., the Hubble constant remains constant in time, it does not change with time. In the Hoyle-Narlikar version of the steady-state cosmological model, known as the C-field (`creation field') theory (Hoyle \& Narlikar 1962) Einstein's field eqs.(1) are modified to
\begin{eqnarray}
R_{\mu\nu} - \frac{1}{2}Rg_{\mu\nu} + C_{\mu\nu} = -\kappa T_{\mu\nu} 
\end{eqnarray}

The C-field is a scalar field with zero mass and zero charge, it represents the phenomenon of creation of matter. In eq.(10) $C_{\mu\nu}$ are second order derivatives of $C$ with respect to the spacetime coordinates $x^\mu$, they define a second rank symmetric tensor field.
\end{enumerate}

Now, if we denote by $dl(t)$ the separation between two neighbouring points $(r,~\theta,~\phi)$ and $(r+dr,~\theta+d\theta,~\phi+d\phi)$ in the 3-space of the spacetime of the universe at the cosmic time t, we see that this separation continually increases with time in proportion to (i)$e^{Ht}$ in the de Sitter model, (ii) S(t) in the Friedmann models, (iii) S(t) in Lema{\^{\i}}tre models, and (iv) $e^{H_\circ t}$ in the steady-state model of the universe. This means that, apart from the material particles, the geometrical points in the 3-space of the spacetime of the universe are also receding away from each other as if they are repelling each other ! This is so in all the relativistic cosmological models of the universe (except in the Einstein universe) during the expansion phase. The pertinent question in this connection is : Why ? However, during the contraction phase in the Friedmann model with k=1 as well as in the Lema{\^{\i}}tre models, the geometrical points in the 3-space of the spacetime of the universe are approaching each other as if they are attracting each other ! Again, the pertinent question is : Why ?

If we denote by $V_{3}(t)$ the volume of the 3-space of the spacetime of the universe at the cosmic time t, we easily infer from the above expressions for the line element that $V_{3}(t)$ continually increases with time in proportion to (i) $e^{3Ht}$ in the de Sitter model (ii) $S^{3}(t)$ in the Friedmann models, (iii) $S^{3}(t)$ in the Lema{\^{\i}}tre models, and (iv) $e^{3H_{\circ}t}$ in the steady-state model during the expansion phase. However, during the contraction phase in the Friedmann model with k=1, and in the Lema{\^{\i}}tre models, $V_{3}(t)$ continually shrinks.

Majority of cosmologists tacitly hold the view that the 3-space of the spacetime of the universe is also expanding apart from the matter and the radiation in it. It is because of this view that continuous creation of matter in the universe at an appropriate rate has been postulated by the proponents of the steady-state model of the universe in order that the density of matter in the universe may remain unchanged with the passage of time in conformity with the perfect cosmological principle.

The Friedmann models of the universe, the bedrock of the so called standard cosmology, are plagued with the problem of singularity apart from that of horizon and flatness. In these models, at some time in the very remote past, conventionally denoted by t=0, the scale factor $S$ was zero; $S(0)=0$. This means, at that epoch, $dl$ and $V_3$ were also zero; $dl(0)=0,~V_3(0)=0$. Consequently, at that epoch not only the entire matter and the radiation in the universe were squeezed into a point in the 3-space (i.e., they occupied zero volume in the 3-space), but also the volume of the 3-space of the universe was zero. In other words at the epoch t=0, the universe was just a point. Therefore, according to these models, the volume of the 3-space of the universe has sprung from zero at the epoch t=0 to its comparatively very large value at the present epoch! In other words, according to these models, even the 'space' was created at the epoch t=0!

\section{There is no observational evidence of the expansion of the 3-space}
However, there is no observational evidence of the expansion of the 3-space of the spacetime of the universe implied in the relativistic cosmological models of the universe. Observation of the recession of galaxies away from each other and that of the thermal cosmic microwave background radiation (CMBR) $-–$ with a very closely Planckian black-body spectrum at T = 2.73K, by Penzics and Wilson (1965) and subsequently by others, whose existence was first predicted by Alpher and Hermann (1948) as a relict of the primordial radiation produced during the nucleosynthesis of lighter elements when the temperature and the density of the matter in the universe were sufficiently high to secure appreciable reaction rates -- have been misconstructed as the evidence of the expansion of the 3-space of the spacetime of the universe.  Observation  of the recession of galaxies away from each other only proves that the matter  in the universe has been continually expanding, it in no way proves that the 3-space of the spacetime of the universe has also been expanding continually. Similarly, the observation of the CMBR only proves that the primordial radiation in the universe has been continually expanding, again it in no way proves that the 3-space of the spacetime of the universe has also been expanding continually.
If at all the 3-space of the spacetime of the universe has also been expanding continually along with the matter and the radiation in the universe, then the following pertinent questions arise:
\begin{enumerate}
\item
One can understand the expansion of the matter and the radiation in the universe by attributing it to the big-bang, i.e. it being triggered off by the big-bang. However, it is incomprehensible how the big-bang can trigger off the expansion of the 3-space of the spacetime of the universe. Consequently, the question is :  What physical process triggered off  the expansion of the 3-space of the universe and what is maintaining its continual expansion ? Certainly, no explosion whatsoever can trigger off the expansion of the 3-space of the spacetime of the universe.

\item
The so called standard model of cosmology implies that before the big-bang the volume of the 3-space of the spacetime of the universe was zero, i.e. the 3-space of the spacetime of the universe was just  a point (a singularity). In other words, the standard model of cosmology implies that the 3-space of the spacetime of the universe has come into existence ex nihilo. Now the question is : Before the big-bang  what was outside the 3-space of zero volume, i.e. outside the singularity, and into what the 3-space started expanding after the big-bang? We know that when a gas expands, the separation between each and every pair of its molecules continually increases; the gas expands into the space surrounding it which already existed before the gas started expanding; the space does not expand at all with the gas. Can the increase in the separation between each and every pair of molecules in the gas be construed as an evidence of expansion of the space into which the gas is embedded? If not, how can the increase in the separation (i.e. the distance) between each and every pair of galaxies ( i.e. the recession of galaxies away from each other) in the universe be construed as the expansion of the 3-space of the spacetime of the universe?

\item
If the 3-space has been continually expanding together with the radiation and the matter in the universe after the big-bang, it implies that the 3-space has been inseparably frozen with (i.e. tagged onto) the matter and the radiation in the universe right from the big-bang epoch like the magnetic lines of force being frozen in the fluid in the case of magnetohydrodynamics. Naturally then, the question aries : What is the physics behind it, i.e. what physical process has caused this freezing of the 3-space of the spacetime of the universe with the matter and the radiation in the universe.

\item
In the radiation era (i.e. during the period between 10 second and $10^{12}$ second after the big-bang) the matter and the radiation were coupled. Was the 3-space also coupled with the matter and the radiation in this era which was dominated by the radiation, and was the 3-spacce expanding with the same speed as the matter and the radiation in this era?

\item
After the recombination epoch (i.e. in the matter era, when the temperature of the primordial radiation had dropped below 3000 K) the matter and the radiation in the universe have decoupled. What about the 3-space? Has it remained coupled with the radiation or with the matter in the universe, or with the  none of the two?

\item
After the recombination epoch (i.e.after the redshift z has been less than 1000) the universe has become transparent to the radiation. Consequently, after this epoch the radiation must have been expanding with the speed of light in vacuum, and the matter must have been expanding with the speed less than that of light in vacuum. What has been the speed of expansion of the 3-space after recombination epoch? Has it been the same as that of the radiation, or the same as that of the matter, or altogether different? And, in any of these case, why?

\item
Earlier, it was believed that the velocity of recession of galaxies is being decelerated. But now there is observational evidence that at the present epoch, it is being accelerated (due to the existence of the ‘dark energy’). What about the velocity of expansion of the 3-space? Is it being decelerated or accelerated ? And, in either case, why? 

\end{enumerate}

To sum up the questions raised above, can \.S( t ), the rate of change of the factor S(t) with time, be taken to be the same for the spacetime, the matter and the radiation in the universe ? If yes, what is the justification from the point of view of physics for this? 

\section{Conclusion} 
The GTR on which the relativistic cosmological models of the universe are based cannot provide unequivocal and satisfactory answers to the questions raised in section 4 above. Moreover, no solution of Einstein's field equation (1) or (2) has, per se, an inflationary phase which has been introduced ad hoc to resolve the horizon problem with which the relativistic cosmological models of the universe are plagued. Furthermore, the GTR does not give any reason as to why the big-bang occurred, if at all it occurred. Actually, the purported expansion of the 3-space of the spacetime of the universe is an inherent feature of the relativistic cosmological models of the universe which has not at all been validated by observation and may not have anything to do with the reality. Consequently, the conceptual foundation of the relativistic cosmological models of the universe based on the GTR, which in turn is based on Riemannian geometry, may be faulty and misleading. However, the Newtonian cosmology (McCrea \& Milne 1934), which is not based on Riemannian geometry and yet 'can describe cosmology in an adequate manner' (Narlikar 2002) does not imply any expansion whatsoever of the space in which galaxies, CMBR, and other celestial entities are embedded.\\

\bigskip
\noindent
{\bf Acknowledgments} \\
The author thanks Professor S. K. Pandey, Co-ordinator of the Reference Centre at Pt. Ravishankar Shukla University Raipur of the Inter-University Centre for Astronomy and Astrophysics, Pune for making available the facilities of the centre. He also thanks Miss Samridhi Kulkarni for typing the manuscript.


\begin{thebibliography}{}
\bibitem{} Alpher R. A.,Hermann R. C., 1948, Nature, 162, 774 
\bibitem{} Bondi H., 1960, Cosmology, Cambridge University Press, p 99
\bibitem{} Bondi H., Gold T., 1948, MNRAS, 108, 252
\bibitem{} de Sitter W., 1917a, Proc. Koninkl. Akad. Weteush, Amsterdam, 19, 1217
\bibitem{} de Sitter W., 1917b, Proc. Koninkl. Akad. Weteush, Amsterdam, 20, 229
\bibitem{} de Sitter W., 1917c, Proc. Koninkl. Akad. Weteush, Amsterdam, 20, 1309
\bibitem{} de Sitter W., 1917d, MNRAS, 78, 3
\bibitem{} Einstein A., 1915, Preuss. Akad. Wiss. Berlin Sitzber. pt. 2, 844 
\bibitem{} Einstein A., 1917, Preuss. Akad. Wiss. Berlin Sitzber. pt. 1, 142
\bibitem{} Einstein A., 1970, in Schilpp P. A., ed., Albert Einstein: Philosopher-Scientist, Open Court Publishing Company, Peru, Illinois, p. 684
\bibitem{} Friedmann A. A., 1922, Z. Phys., 10, 377
\bibitem{} Friedmann A. A., 1924, Z. Phys., 21, 326
\bibitem{} Gamow G., 1970, My World Line Viking, New York, p. 44
\bibitem{} Hoyle F., 1948, MNRAS, 108, 372
\bibitem{} Hoyle F., Narlikar J. V., 1962, Proc. Roy. Soc., A270, 334
\bibitem{} Hubble E. P., 1929, Proc. Natl. Acad. Sci, U.S.A., 15, 168
\bibitem{} Lema$\hat{\imath}$tre G., 1927, Ann. Soc. Sci., Brux., A47, 49
\bibitem{} Lema$\hat{\imath}$tre G., 1931, MNRAS, 91, 483
\bibitem{} Lema$\hat{\imath}$tre G., 1970, Schilpp P. A., ed. in Albert Einstein: Philosopher-Scientist, Open Court Publishing Company, Peru, Illinois, p. 437
\bibitem{} Lundmark K., 1920, Stock. Acad. Hand., 50, No. 8
\bibitem{} Lundmark K., 1924, MNRAS, 84,747
\bibitem{} Lundmark K., 1925, MNRAS, 85, 865
\bibitem{} McCrea W. H., Milne E. A., 1934, Q. J. Math. 5, 73
\bibitem{} Narlikar Jayant Vishnu, 2002, An Introduction to cosmology, Cambridge University Press, Cambridge, p. 36
\bibitem{} Peebles P. J. E.,  Ratra Bharat, 2003, Rev. Mod.Phys., 75, 559
\bibitem{} Penzias A. A., Wilson R. W., 1965, ApJ, 142, 419
\bibitem{} Robertson H. P., 1935, ApJ, 82, 248 
\bibitem{} Slipher V. M., 1914, Paper presented at the 17th meeting of AAS
\bibitem{} Slipher V. M., 1924, in Eddington A. S., The Mathematical Theory of Relativity, Cambridge University Press, Cambridge, p. 162
\bibitem{} Walker A. G., 1936, Proc. Lond. Math. Soc. (2), 42, 90 
\bibitem{} Wirtz C., 1921, Astr. Nachr., 215, 349
\bibitem{} Wirtz C., 1922, Astr. Nachr., 216, 451
\bibitem{} Wirtz C., 1924, Astr. Nachr., 222, 21 
\bibitem{} Wirtz C., 1925, Scientia, 38, 303

\end{thebibliography}
\end{document}